\documentclass[aps,twocolumn,preprintnumbers,amsmath,amssymb]{revtex4}
\usepackage{amssymb}
\usepackage{graphicx,color}

\newcommand{\ber}{\begin{eqnarray}}
\newcommand{\eer}{\end{eqnarray}}
\newcommand{\snn} {\sqrt{s_{_{\rm NN}}}}
\newcommand{\tauf} {\tau_{_{\rm F}}}
\newcommand{\detr} {dE_{_{\rm T}}}
\newcommand{\dtwoetr} {d^2E_{_{\rm T}}}
\newcommand{\area}{A_{_{\rm T}}}
\newcommand{\dt}{d_t}
\newcommand{\lefttwo}{\!\!}
\newcommand{\leftfive}{\!\!\!\!\!\!}
\newcommand{\ycm}{y_{_{\rm CM}}}
\newcommand{\np}{N_{\rm part}}
\newcommand{\tone}{t_1}
\newcommand{\ttwo}{t_2} 
\newcommand{\ttwoone}{t_{21}}
\newcommand{\tmid}{t_{mid}}
\newcommand{\emax}{\epsilon^{max}}
\newcommand{\ebj}{\epsilon_{_{\rm Bj}}}

\begin{document}
\title{Extension of the Bjorken energy density formula of the initial
  state for relativistic heavy ion collisions}
\author{Zi-Wei Lin}
\address{Key Laboratory of Quarks and Lepton Physics (MOE) and
  Institute of Particle Physics, Central China Normal University, Wuhan 430079, China}
\address{Department of Physics, East Carolina University, Greenville, NC 27858, USA}
\email{linz@ecu.edu}
\date{\today}

\begin{abstract}
For relativistic heavy ion collisions, the Bjorken formula is very 
useful for estimating the initial energy density once an
initial time $\tau_0$ is specified. However, it cannot be trusted at low energies,
e.g. well below $\sqrt{s_{_{\rm NN}}} \approx 50$ GeV for central Au+Au
collisions, when $\tau_0$ is smaller than the finite time it takes for
the two nuclei to cross each other. Here I extend the Bjorken formula
by including the finite time duration of the initial energy
production. Analytical solutions for the formed energy density in the
central spacetime-rapidity region are derived for several time profiles. 
Compared to the Bjorken formula at low energies, the maximum energy
density reached is much lower, increases much faster with the
collision energy, and is much less sensitive to the uncertainty of the
formation time, while the energy density time evolution is much longer. 
Comparisons with results from a multi-phase transport  
confirm the key features of these solutions. The effect of the
finite longitudinal width of the initial energy production, which is
neglected in the analytical results, is investigated with the
transport model and shown to be small. 
This work thus provides a general model for the initial energy production of
relativistic heavy ion collisions that is also valid at low energies.

\end{abstract}
\maketitle

\section{Introduction}
Relativistic heavy ion collisions aim to create the quark-gluon 
plasma (QGP) and study its properties \cite{Adams:2005dq,Adcox:2004mh}. 
Therefore it is important to better understand the initial energy
production, including the maximum value and time evolution of the energy
density in the overlap volume.  
For low energies such as the Beam Energy Scan at the Relativistic
Heavy Ion Collider, the relationship between the time evolution of the
energy density or net-baryon density and the possible critical point
of QCD becomes important \cite{Stephanov:2011pb,Bialas:2016epd}.
The Bjorken formula \cite{Bjorken:1982qr}
is a very useful tool in estimating the initial energy density in the
central rapidity region after the two nuclei pass each other:
\ber
\ebj(t)=\frac{1}{\area \; t} \frac{\detr}{dy}.
\label{enebj}
\eer
In the above, 
$\area$ represents the full transverse area of the overlap volume, 
and $\detr/dy$ is the rapidity density of the transverse
energy at mid-rapidity (at an early time $t$), which is often
approximated with the experimental $\detr/dy$ value in the final state. 
Since the Bjorken energy density diverges as $t \rightarrow 0$,
a finite value is needed for the initial time $\tau_0$ \cite{Adcox:2004mh}.
Considering that the production of a particle takes a finite formation
time $\tauf$, one can take the Bjorken formula at time $\tauf$ to
obtain the initial formed energy density.

A severe limitation of the above Bjorken energy density formula of the
initial state results from the fact that it neglects the finite
thickness of the colliding nuclei (along the beam direction $z$),
which leads to a finite duration time, as well as  
a finite longitudinal width in $z$, for the initial energy production. 
Using a hard-sphere model for the nucleus, it will take the following
time for two nuclei of the same mass number $A$ to cross each other in a 
central collision in the center-of-mass frame \cite{Kajantie:1983ia}:
\ber
\dt=\frac{2 R_A}{\sinh \ycm},
\eer
where $\ycm$ is the projectile rapidity in the center-of-mass frame
and $R_A$ is the nuclear radius. 
Therefore the Bjorken formula is only valid when the duration time 
(or crossing time) is much smaller than the formation time $\tauf$ 
\cite{Adcox:2004mh}.
As an example, for $\tau_0=0.5$ fm/c, the Bjorken
formula cannot be trusted for central Au+Au collisions well below
$\snn \approx 50$ GeV since $\dt \approx 0.5$ fm/c there.

My goal of this study is to derive a Bjorken-like formula so
that it is also valid at low energies where the Bjorken formula
breaks down. I accomplish this by including the finite crossing time 
in the time profile of the initial energy production. 
I focus on the formed energy density, 
averaged over the full transverse overlap area, in the central
spacetime-rapidity region ($\eta_s=0$) in the center-of-mass frame of
central collisions of two identical nuclei.

\section{Method}

\begin{figure}[h]
\includegraphics[width=3.35 in]{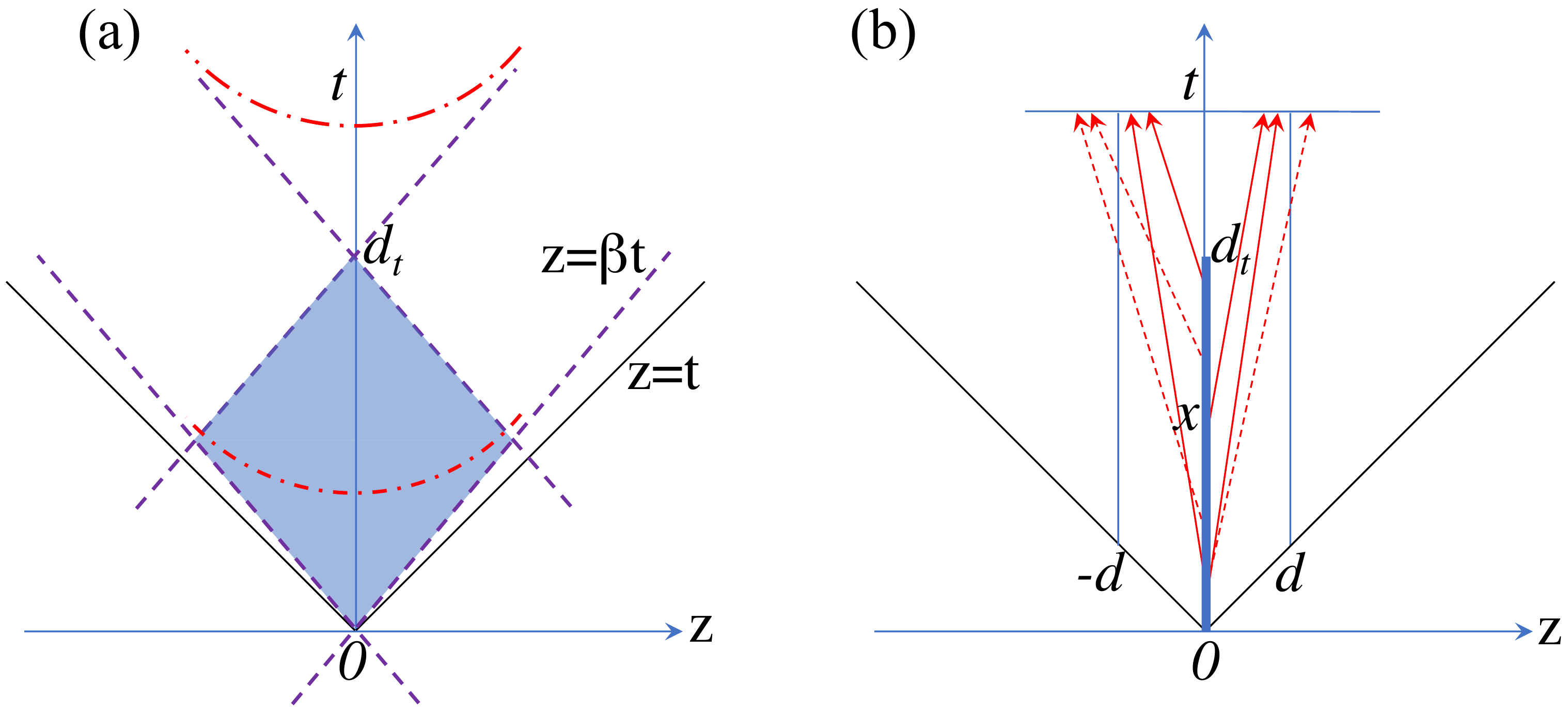}
\caption{(a) Schematic picture of the collision of two nuclei, 
where the initial energy production takes place over the shaded area
of finite widths in $t$ and $z$.
(b) This simplified picture is considered for analytical derivations of
energy density at central spacetime-rapidity $\eta_s \rightarrow 0$ (as $d
\rightarrow 0$): particles could be initially produced at any time $x$ within $[0,
\dt]$ at $z=0$ and then propagate to observation time $t$. 
} 
\label{fig:drawing}
\end{figure}

Since the Bjorken formula \cite{Bjorken:1982qr} 
is only valid at very high energies where the two incoming nuclei are
highly Lorentz-contracted \cite{Adcox:2004mh}, it essentially assumes
that the initial energy production occurs at time $t=0$. Then the
quanta appear (i.e. are formed) after a certain proper time $\tau_0$,
as shown by the lower dot-dashed hyperbola in Fig.~\ref{fig:drawing}(a). 
This proper time can be viewed as a typical decay time of the color
fields created from primary collisions of the two nuclei
\cite{collection}. Because of $y=\eta_s$ in the Bjorken ansatz, 
the quanta appearing at $\eta_s=0$ or $y=0$ are initially 
produced at time $t=0$ on the $z=0$ plane. 

Once one considers the finite crossing time of the two nuclei,
however, the initial energy production actually goes on throughout
this period of time.
Figure~\ref{fig:drawing}(a) shows a schematic picture, where 
the two nuclei come into contact at time $0$ and pass each other at time
$\dt$ \cite{Kajantie:1983ia}. 
The two solid diagonal lines represent the light-cone boundaries
(in natural units where the speed of light has been set to one),
while each pair of the parallel dashed lines represents the
boundaries of the $t-z$ trajectories of nucleons in an incoming
nucleus that moves with speed $\beta$. 
The shaded area, indicating the primary nucleon-nucleon
collision region, shows that the initial energy production 
takes place over a finite amount of time. 
Again assuming a proper formation time $\tau_0$, 
primary collisions at time $0$ will then produce formed quanta on the lower
hyperbola while primary collisions at time $\dt$ will produce formed
quanta on the upper hyperbola as shown in Fig.~\ref{fig:drawing}(a). 
In addition, one sees that the initial energy is produced over a finite
longitudinal width; for example, a particle initially produced at $z >
0$ may propagate with a negative rapidity and later cross the $z=0$ plane.

To obtain analytical results for the central spacetime-rapidity
region, I include the finite duration time
in my formulation but neglect the finite longitudinal $z$-width in
the initial energy production. 
Figure~\ref{fig:drawing}(b) shows the simplified schematic
picture for the region at $\eta_s \approx 0$, where the initial
particles and energy are assumed to be produced over the crossing time
but at $z=0$. In order to obtain analytical results, I make minimal 
extensions to the Bjorken formula framework 
\cite{Bjorken:1982qr}, thus I also 
neglect secondary particle interactions or the transverse expansion. 
The numerical results from a multi-phase transport (AMPT) model
\cite{Lin:2004en}, however, include the
finite longitudinal width in $z$, secondary parton scatterings, and
the transverse expansion. In particular, 
I shall study with AMPT the effect of the
finite longitudinal width in the initial energy production, which is
neglected in the analytical formulation.
Note that I only address the Bjorken energy density formula of the
initial state \cite{Bjorken:1982qr,Adcox:2004mh} shown as
Eq.(\ref{enebj}), not the more general Bjorken model 
\cite{Bjorken:1982qr} that also 
assumes local thermal equilibrium for the produced quanta in the
initial state and then considers the subsequent hydrodynamic
space-time evolution.
Furthermore, since I only consider the central spacetime-rapidity
region, I have written the time variable as $t$ instead of $\tau$. 

Let us write the production rate of the initial
transverse energy rapidity density 
around $y \approx 0$ at time $x$ as $\dtwoetr/dy/dx$.
Thus there could be particle productions at any time $x \in
[0, \dt]$, while $\dtwoetr/dy/dx=0$ for $x<0$ or $x>\dt$. 
With the picture of Fig.~\ref{fig:drawing}(b), I 
evaluate the energy density within a narrow region $-d \leq z
\leq d$ at time $t > \dt$. For a particle produced at time $x$ to
stay within this $z$-region, its rapidity needs to
satisfy 
\ber
|\tanh y| \approx |y| \leq \frac{d}{t-x} 
\eer 
at $y \approx 0$. Note that the right-hand-side above can always be made small with
small-enough $d$, so that $\dtwoetr/dy/dx$ does not depend on $y$
within this small $y$-range. 
Therefore the average energy density in this region at time $t$ is
\ber
\frac{E}{2d \area}
=\frac{1}{\area} \int_0^{\dt} \frac{\dtwoetr}{dy\;dx} \frac{dx}{(t-x)}.
\eer

From now on I shall study the formed energy density by assuming a
finite formation time $\tauf$ for the produced particles. A similar
analysis gives the following average formed energy density at
any time $t \geq \tauf$ as
\ber
\epsilon(t)=\frac{1}{\area} \int_0^{t-\tauf} \frac{\dtwoetr}{dy\;dx}
\frac{dx}{(t-x)}.
\label{egeneral}
\eer
As in the Bjorken formula, $\epsilon (t<\tauf)=0$.
However, an important feature of the above formula is that it 
applies to early times when the two nuclei are still crossing each
other (i.e. $t \leq \dt+\tauf$).
Note that Eq.(\ref{egeneral}) above reduces to the Bjorken formula
when one neglects the finite crossing time by taking 
$\dtwoetr/dy/dx \rightarrow \delta (x)~\detr/dy$. 
To proceed further, I will next take specific forms for the time
profile of the initial energy production $\dtwoetr/dy/dx$.

\section{Results}
\begin{figure}[h]
\includegraphics[width=3.3 in]{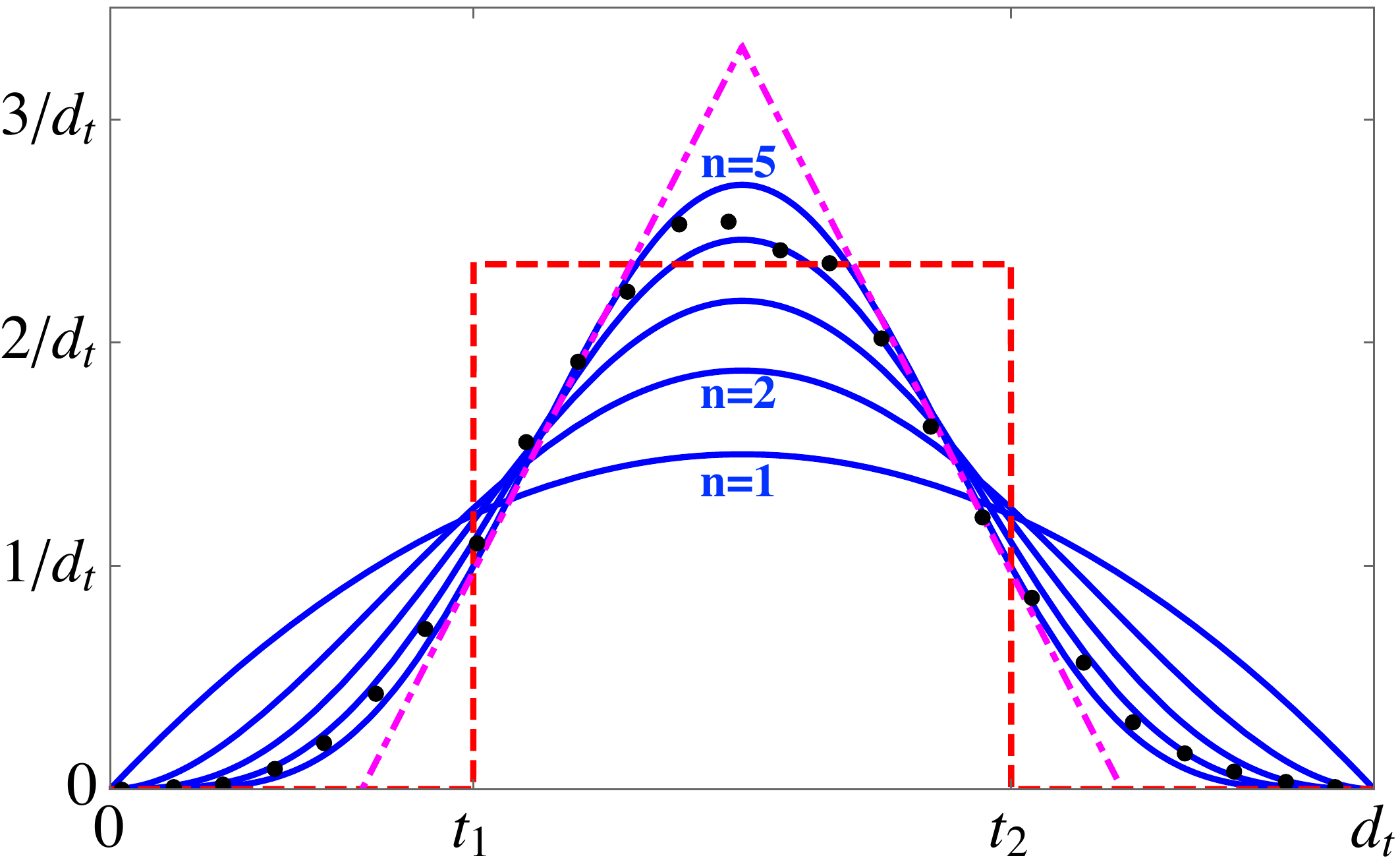}
\caption{Time profiles for the initial energy production at
  central spacetime-rapidity: a uniform profile (dashed curve), beta
  profiles with integer powers $n=1$ to $5$ (solid curves), and  a
  triangular profile (dot-dashed). Circles represent the time profile of
  partons within mid-spacetime-rapidity from the string melting AMPT model for
  central Au+Au collisions at $\snn=11.5$ GeV.
} 
\label{fig:tprofile}
\end{figure}

For simplicity, I first assume that the initial energy is 
produced uniformly from time $\tone$ to $\ttwo$ (with $\ttwoone \equiv
\ttwo-\tone$): 
\ber
\frac{\dtwoetr}{dy\;dx}= {1 \over \ttwoone} \frac{\detr}{dy},
{~\rm if~} x \in [\tone,\ttwo].
\label{uniform}
\eer
Note that one only needs the above assumption to apply at
$y \approx 0$.  Also, I have not related 
$\tone$ and $\ttwo$ to $\dt$ for the sake of generality. 
An illustration of this time 
profile is shown as the dashed curve in Fig.~\ref{fig:tprofile}.
Equation~(\ref{egeneral}) then gives the following
solution for the formed energy density:
\ber
\epsilon_{\rm uni}(t)&& \leftfive=
\frac{1}{\area \ttwoone} \frac{\detr}{dy} \ln \!
\left ( \! \frac{t-\tone}{\tauf} \! \right ), \lefttwo {~\rm if~} 
t \in [\tone+\tauf,\ttwo+\tauf]; \nonumber \\
&&\leftfive =\frac{1}{\area \ttwoone} \frac{\detr}{dy} \ln \!
\left ( \! \frac{t-\tone}{t-\ttwo} \! \right ),  \lefttwo {~\rm if~} 
t \geq \ttwo+\tauf.
\label{solnuni}
\eer
One can easily verify that, for $\tone=0$ and $\ttwo/\tauf \rightarrow 0$, this 
solution reduces to the Bjorken formula of Eq.(\ref{enebj}).

Qualitatively, this energy density starts from 0 at time
$\tone+\tauf$, grows smoothly to the following maximum value $\emax$ at time
$\ttwo+\tauf$, and then decreases abruptly after the energy production
stops: 
\ber
\emax_{\rm uni} = \epsilon_{\rm uni} (\ttwo+\tauf)=
\frac{1}{\area \ttwoone} \frac{\detr}{dy} \ln \!
\left ( \! 1+\frac{\ttwoone}{\tauf} \! \right ).
\label{emaxuni}
\eer
Compared to the maximum energy density $\ebj \!(\tauf)$ given by the
Bjorken formula, one has
\ber
\frac{\emax_{\rm uni}}{\ebj (\tauf)} = 
\frac{\tauf}{\ttwoone} \ln \! \left ( \! 1+\frac{\ttwoone}{\tauf} \!
\right ).
\label{euni}
\eer
Therefore the $\emax$ value above is always smaller than the 
Bjorken initial energy density:
$\emax \ll \ebj \!(\tauf)$ at low energies where $\tauf/\ttwoone$
is small, while at high energies $\emax \approx \ebj \!(\tauf)$. 
Furthermore, as $\tauf/\ttwoone \rightarrow 0$, the peak energy
density $\emax$ grows as $\ln (1/\tauf)$, much slower
than the $1/\tauf$ growth of the Bjorken formula. This means that,
after taking into account the finite crossing time, the maximum energy
density achieved will be much less sensitive to the uncertainty of
$\tauf$, especially at lower energies where $\ttwoone 
\approx \mathcal{O}(\dt)$ is bigger.  
In addition, Eq.(\ref{solnuni}) shows that the initial energy density
at time later than $\ttwo+\tauf$ is independent of $\tauf$. One will
see next that these features are general and also apply
to the other time profiles. 

Due to the typical spherical shape of a nucleus, there will be 
few primary nucleon-nucleon interactions when the two nuclei barely
touch or almost pass each other, while there will be many
such interactions when the two nuclei fully overlap (around time
$\dt/2$). I thus expect the time profile of the initial energy
production to peak around time $\dt/2$ while diminish at time 0 and
$\dt$. Therefore I can choose the following time profile based on 
the probability density function of the beta distribution with equal
shape parameters:
\ber
\frac{\dtwoetr}{dy\;dx} = a_n \left [ x (\dt-x) \right ]^n
\frac{\detr}{dy},
{~\rm if~} x \in [0,\dt].
\label{beta}
\eer
In the above, power $n$ does not need to be an integer, 
and $a_n=1/\dt^{2n+1}/B(n+1,n+1)$ is the
normalization factor with $B(a,b)$ being the Beta function.
This smooth beta profile reduces to a uniform
profile when $n=0$; with an appropriate value of $n$ it can also well
describe the transport model time profile, as shown in 
Fig.~\ref{fig:tprofile}.
I obtain the following solution for the formed energy density:
\ber
\epsilon_{\rm beta} (t)&& \leftfive=
\frac{1}{\area} \frac{\detr}{dy} 
\frac{\left [(t-\tauf)/\dt \right ]^{n+1}}{(n+1) B(n+1,n+1) \;t} \nonumber \\ 
&& * F_1 \lefttwo \left
  [n+1,-n,1,n+2,\frac{t-\tauf}{\dt},\frac{t-\tauf}{t} \right ],
\nonumber \\  
&& {~\rm if~} t \in [\tauf,\dt+\tauf]; \nonumber \\
&& \leftfive =\frac{1}{\area} \frac{\detr}{dy} \frac{1}{t}
*_2\lefttwo F_1 \lefttwo \left [ 1,n+1,2n+2,\frac{\dt}{t} \right ],
\nonumber \\
&& {~\rm if~} t \geq \dt+\tauf. 
\eer
$F_1$ above is the Appell hypergeometric function of
two variables, and $_2F_1$ is the Gaussian hypergeometric function. 
One can verify that for $n=0$ the above solution reduces to
Eq.(\ref{solnuni}) for $\tone=0$ \& $\ttwo=\dt$. 

I now apply these solutions to central Au+Au collisions. 
The nuclear transverse area is taken as 
\ber
\area=\pi R_A^2, {\rm~with~} R_A=1.12A^{1/3} {~\rm fm},
\eer
where $A=197$. I take the mid-rapidity $\detr/dy$ as the following
data-based parameterization \cite{Adler:2004zn}:
\ber
\frac{\detr}{dy}=1.25 \frac{\detr}{d\eta}=0.456~\np 
\ln \lefttwo \left ( \! \frac{\snn}{2.35} \right ),
\label{detdy}
\eer 
where $\snn$ must be greater than $2.35$ in the
unit of GeV. Also, I take $\np=2A$ for central collisions.

\begin{figure}[htb]
\includegraphics[width=3.3 in]{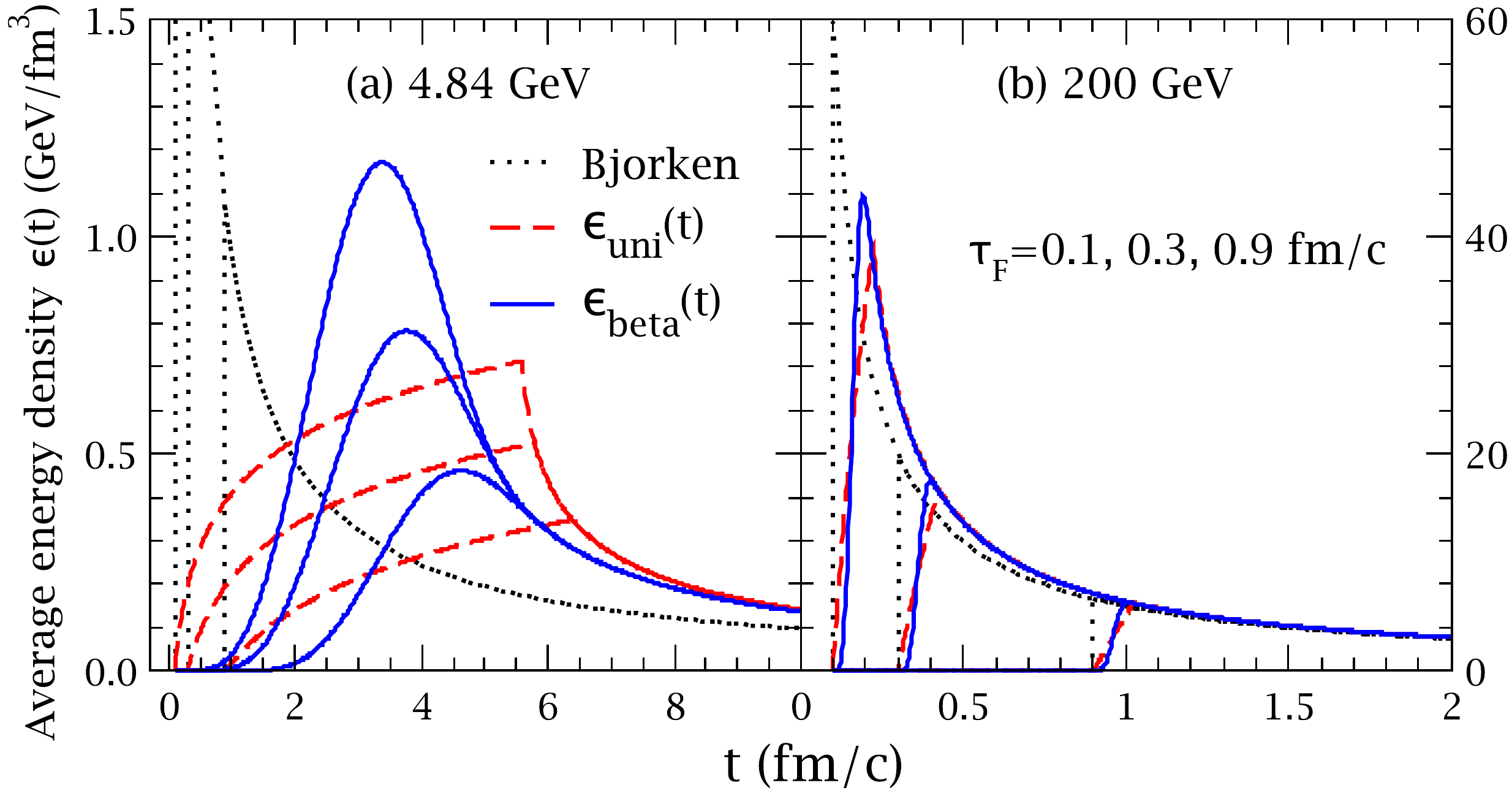}
\caption{
Average formed energy densities at central spacetime-rapidity as functions of time for
central Au+Au collisions at (a) 4.84 GeV and (b) 200 GeV from the
uniform time profile  with the naive choice of $\tone=0$ \&
$\ttwo=\dt$ (dashed), the beta time profile for $n=4$ (solid), and the Bjorken
formula (dotted). Three sets of curves of each type correspond to
$\tauf=0.1,0.3$ \& $0.9$ fm/c. 
} 
\label{fig:eden}
\end{figure}

My results for central Au+Au collisions at $\snn=4.84$ GeV and 200
GeV are shown in Fig.~\ref{fig:eden} for different formation times
$\tauf=0.1,0.3$ \& $0.9$ fm/c. Also shown are the results implied by
the Bjorken formula: $\ebj(t)=1/(\area t) \detr/dy$ for $t \geq \tauf$
(and $=0$ for $t < \tauf$).
I have taken $n=4$ for the beta profile according to
Fig.~\ref{fig:tprofile}, 
and I take the naive choice of $\tone=0$ and $\ttwo=\dt$ for the
uniform profile in Fig.~\ref{fig:eden}. 
At 4.84 GeV, one sees that the time evolution of the energy density in either time
profile has a much bigger width (e.g. full width at half maximum) than
the Bjorken results, while the maximum energy density is much lower
than the corresponding Bjorken value for the same $\tauf$. 
As expected, my maximum initial energy density $\emax$
changes by a much smaller factor of 2.1 (uniform profile) 
or 2.5 (beta profile) 
when $\tauf$ changes from $0.1$ to $0.9$ fm/c; 
while the Bjorken initial energy density changes by a factor of 9.
On the other hand, my results at 200 GeV are much closer to (although
still different from) the Bjorken results; this is expected since the
crossing time there ($\dt \approx 0.12$ fm/c) is very small. 
For both energies, my results approach the Bjorken results at late times.

Both the Bjorken formula and my method have neglected secondary
particle interactions and the transverse expansion, which could 
affect the time evolution of the energy density. 
These dynamics can be described by transport models such as AMPT
\cite{Lin:2004en} or hydrodynamic models
\cite{Shen:2012vn,Oliinychenko:2014tqa}. Now I compare my 
analytical solutions with results from the string melting AMPT model,
which includes a conversion of excited strings into a parton matter,
partonic scatterings, a quark coalescence for hadronization, and
hadronic scatterings. For this study, the string melting AMPT model
\cite{Lin:2004en} has been improved by including the finite thickness
of nuclei \cite{work}, then I calculate the average local energy
density (over the hard-sphere transverse area $\area$)  
for partons at mid-spacetime-rapidity following the method of an
earlier study \cite{Lin:2014tya}. 
Circles in Fig.~\ref{fig:tprofile} represent the distribution of
production time of partons within 
mid-spacetime-rapidity from AMPT for central ($b=0$ fm) Au+Au
collisions at $\snn=11.5$ GeV \cite{work}. 
I thus take $n=4$ for the beta time profile, since this can 
reasonably describe the AMPT time profile. 
To get the same mean and standard deviation as the beta profile (for
$n=4$), I set $\tone=0.29 \dt$ \& $\ttwo=0.71 \dt$ for the uniform
profile.

\begin{figure}[htb]
\includegraphics[width=3.3 in]{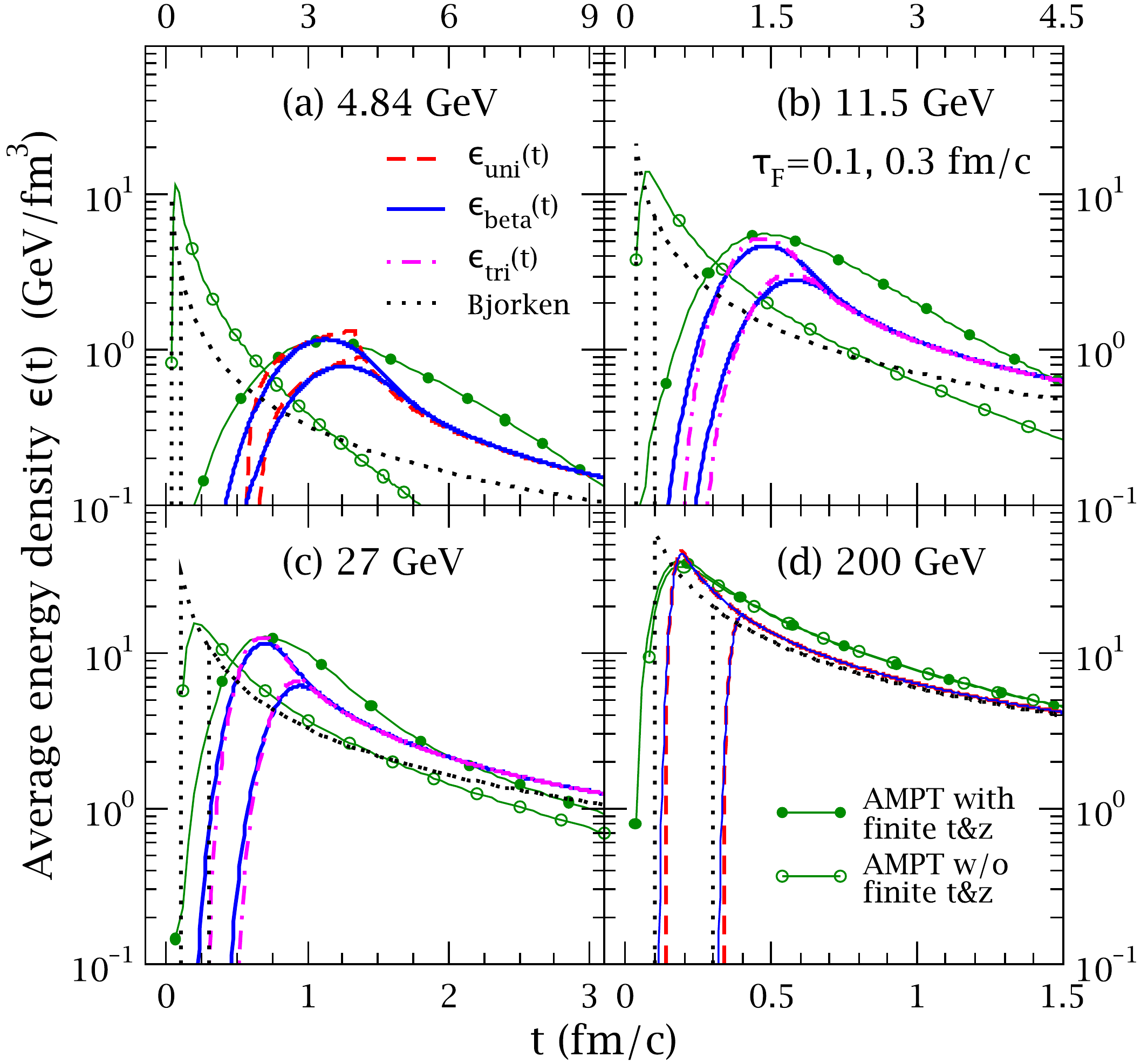}
\caption{
Average energy densities at central spacetime-rapidity for the uniform (dashed curves), beta
(solid curves), triangular (dot-dashed curves) time profiles and the
Bjorken formula (dotted curves) for $\tauf=0.1$ \& $0.3$ fm/c, in
comparison with the corresponding AMPT results (circles), for central
Au+Au collisions at (a) 4.84 GeV, (b) 11.5 GeV, (c) 27 GeV, and (d)
200 GeV. I have used $\tone=0.29 \dt$ \& $\ttwo=0.71 \dt$ for the uniform
profile and $\tone=0.20 \dt$ \& $\ttwo=0.80 \dt$ for the triangular
profile so that they both have the same mean and standard deviation as the beta profile. 
}
\label{fig:ampt}
\end{figure}

Figure~\ref{fig:ampt} shows my results from different time
profiles together with the Bjorken results at different energies. 
One sees from Figs.~\ref{fig:ampt}(a)\&(d) that,  
unlike Figs.~\ref{fig:eden}, results from the uniform and beta
profiles here are quite close to each
other once the uniform profile is set to the same mean and standard
deviation as the beta profile.
Curves with filled and open circles are respectively the AMPT
results with and without the finite nuclear thickness.
Note that each AMPT curve with finite thickness 
has been shifted a bit in time, so that it peaks at the same time as
the corresponding beta profile for $\tauf=0.1$ fm/c, in order to better compare their
shapes.
One sees that at the high energy of 200 GeV the AMPT results with and
without the finite nuclear thickness are essentially the same; 
the Bjorken result and my analytical results are also very
similar (especially after allowing shifts in time). This confirms that
expectation that the finite nuclear thickness can be mostly neglected
at high-enough energies. 
One also sees that the AMPT results are generally wider in time; partly
because the parton proper formation time in AMPT is not set as a constant but
is inversely proportional to the parent hadron transverse mass
\cite{Lin:2004en}; I find that the parton formation time distribution 
at mid-spacetime-rapidity has a mean value of $\approx 0.3$ fm/c but 
has a long tail. 
The finite $z$-width for the initial energy production, 
secondary parton scatterings, the transverse
expansion, and the effective work done during the expansion
\cite{Gyulassy:1983ub,Ruuskanen:1984wv} of the dense matter in AMPT
can also cause differences from the analytical results. 
Overall, one sees that the AMPT results without considering the finite
nuclear thickness are similar to the Bjorken results, while the AMPT
results including the finite thickness are similar to my analytical
results. 

\section{Discussions}
One can also take a triangular time profile, as illustrated by the
dot-dashed curve in Fig.~\ref{fig:tprofile}, 
from time $\tone$ to $\ttwo$ with the peak at $\tmid \equiv (\tone+\ttwo)/2$:
$\dtwoetr/dy/dx \propto (x-\tone)$ for $t \in [\tone,\tmid]$ while 
$\propto (\ttwo-x)$ for $t \in [\tmid,\ttwo]$.
One then obtains the following solution:
\ber
\epsilon_{\rm tri} (t)&& \leftfive=
\frac{4}{\area \ttwoone^2} \frac{\detr}{dy} \lefttwo
\left [ -t+\tone+\tauf + (t-\tone) \ln \! \left (\! \frac{t-\tone}{\tauf}
 \! \right ) \right ] \!, \nonumber \\ 
&&{~\rm if~} t \in [\tone+\tauf,\tmid+\tauf]; \nonumber \\ 
&&\leftfive=\frac{4}{\area \ttwoone^2} \frac{\detr}{dy} \lefttwo
\left [ t-\ttwo-\tauf +(t-\tone) \ln \! \left ( \!
    \frac{t-\tone}{t-\tmid} \! \right ) \right . \nonumber \\ 
&& \left . + (\ttwo-t) \ln \! \left ( \! \frac{t-\tmid}{\tauf}
    \! \right ) \right ] \!, \lefttwo {~\rm if~} t \in [\tmid \!+\!
\tauf,\ttwo \!+\! \tauf]; \nonumber \\ 
&&\leftfive=\frac{4}{\area \ttwoone^2} \frac{\detr}{dy} \lefttwo
\left [ (t-\tone) \ln \! \left ( \! \frac{t-\tone}{t-\tmid} \! \right )  \right . \nonumber \\ 
&& \left . +(\ttwo-t) \ln \! \left ( \! \frac{t-\tmid}{t-\ttwo} \!
  \right ) \right ] \!, {~\rm if~} t \geq \ttwo+\tauf.
\eer
This energy density increases smoothly to the
following maximum value $\emax$ at a time within 
$(\tmid+\tauf,\ttwo+\tauf)$ and then decreases smoothly with time:
\ber
\emax_{\rm tri} && \leftfive= \epsilon_{\rm tri} \! \left ((\tone+\ttwo+\tauf+\sqrt {\tauf}
\sqrt {2\;\ttwoone+\tauf}\;)/2 \right ) \nonumber \\
&&\leftfive=\frac{2}{\area \ttwoone} \frac{\detr}{dy} \left [
\frac{}{}
  -1-\frac{\tauf}{\ttwoone}+\sqrt {\frac{\tauf}{\ttwoone}} \sqrt
  {2 +\frac{\tauf}{\ttwoone}} \right . \nonumber \\  
&& \left . +2 \ln \lefttwo \left ( \frac {1+\sqrt
      {1+2\;\ttwoone/\tauf}}{2} \right ) \right ]. 
\label{emaxtri}
\eer

Figures~\ref{fig:ampt}(b)\&(c) show that
results from the beta and triangular profiles are almost identical in
shape and close in magnitudes, 
after I set $\tone=0.20 \dt$ \& $\ttwo=0.80 \dt$ for the 
triangular time profile to have the same mean and standard deviation
as the beta profile for $n=4$. 
An advantage of the triangular profile is that one has analytical
expressions for its $\emax$ and the corresponding time. 

\begin{figure}[htb]
\includegraphics[width=3.3 in]{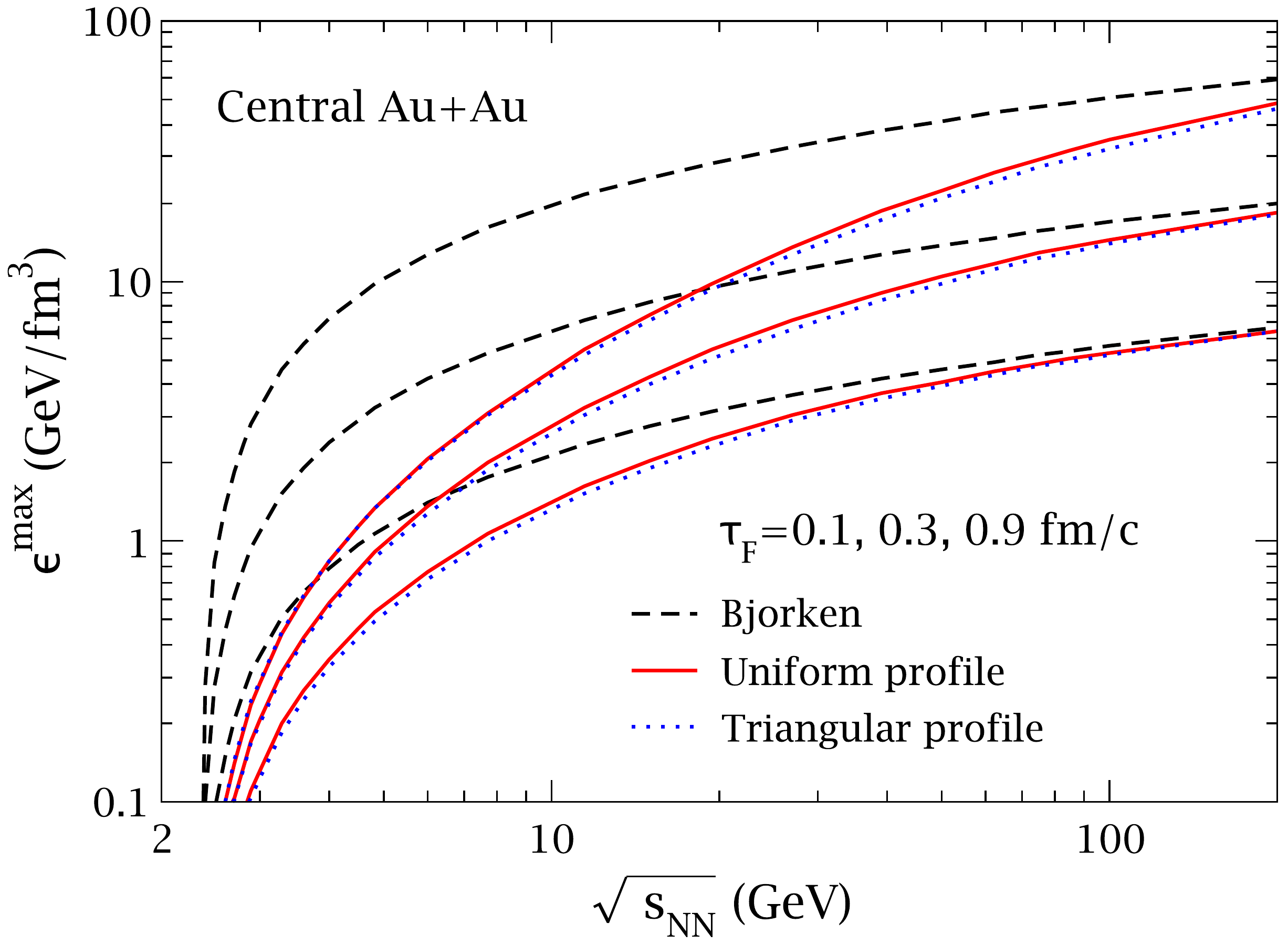}
\caption{Maximum energy density at central spacetime-rapidity
  ($\eta_s=0$) averaged over the transverse overlap area versus the
  collision energy for central Au+Au collisions from the Bjorken
  formula (dashed curves), the uniform profile (solid curves), and the
  triangular profile (dotted curves) for $\tauf=0.1, 0.3$ \& $0.9$
  fm/c. I have used $\tone=0.29 \dt$ \& $\ttwo=0.71 \dt$ for the
  uniform profile and $\tone=0.20 \dt$ \& $\ttwo=0.80 \dt$ for the
  triangular profile. 
} 
\label{fig:emax}
\end{figure}

Figure~\ref{fig:emax} compares $\emax$, the maximum value of the average energy
density at central spacetime-rapidity $\eta_s=0$, in central Au+Au
collisions from the Bjorken formula with that from my analytical
extension, including the uniform and triangular time profiles that
have analytical solutions for $\emax$. 
My results from the uniform and triangular time profiles are quite
close to each other after their $\tone$ and $\ttwo$ parameters are
chosen so that each profile has the same mean and standard deviation
as the beta profile for $n=4$. 
One sees that the increase of the maximum
energy density with the collision energy is much faster than the
prediction from the Bjorken formula; this is consistent with
Eq.(\ref{euni}), which shows that the Bjorken formula overestimates the
maximum energy density more at lower energies.
The overestimation of $\emax$ by the Bjorken formula is also more severe
for smaller $\tauf$. At high energies, however, one sees that my results approach 
the Bjorken formula at the same $\tauf$. 
Note that these results are obtained using the $\detr/dy$
parameterization in Eq.(\ref{detdy}) \cite{Adler:2004zn}, 
which precision should be improved at very low collision energies, 
e.g. when $\snn < 3$ GeV.

\begin{figure}[htb]
\includegraphics[width=3.3 in]{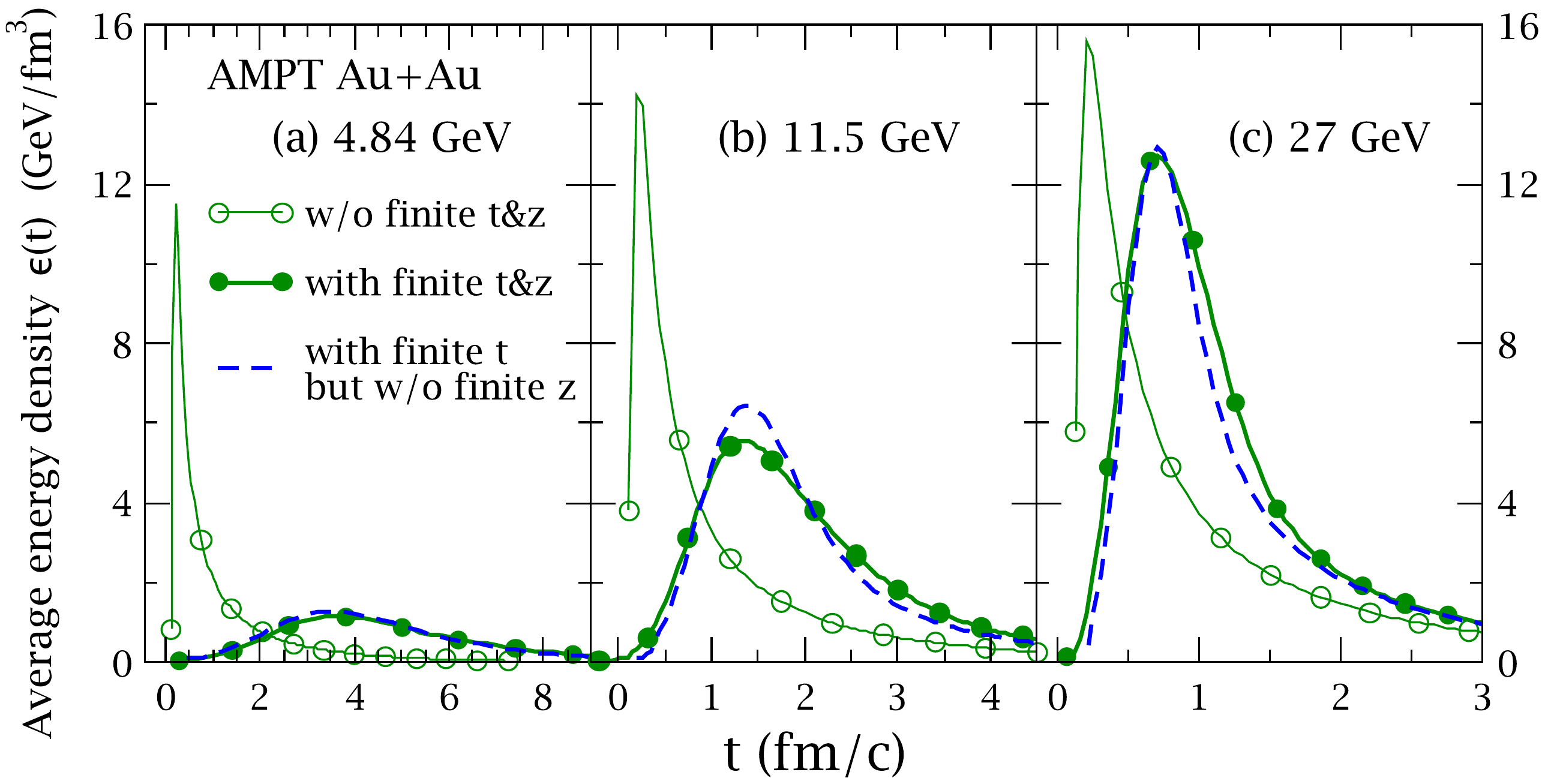}
\caption{AMPT results of average energy densities at central spacetime-rapidity for central
  Au+Au collisions at (a) 4.84 GeV, (b) 11.5 GeV, and (c) 27 GeV when
  excluding the finite widths in $t$ and $z$ (open circles), including
  the finite widths in $t$ and $z$ 
  (filled circles), and including the finite width in $t$ but not the
  finite width in $z$ (dashed curves). 
} 
\label{fig:ampt0z}
\end{figure}

Since my analytical method includes the finite time duration but neglects
the finite $z$-width for the initial energy production, further work
may be warranted to include this effect analytically. 
Note that the finite width in $z$ is already included in one set of
the AMPT results (curves with filled circles) shown in Fig.~\ref{fig:ampt}. 
Here I further demonstrate this effect numerically in
Fig.~\ref{fig:ampt0z}. By modifying the AMPT model to 
include the finite width in $t$ but not the finite width in $z$, 
I obtain the dashed curves in Fig.~\ref{fig:ampt0z}; they are quite 
close to the corresponding full AMPT result (filled circles) in both the peak
magnitude and shape (with the width slightly smaller), while at low
energies they are
very different from the AMPT results that neglect both finite widths
in $t$ and $z$ (open circles).
These results suggest that the effect of 
the finite width in $z$ on my analytical results is rather small.
Again note that, in order to better compare the shapes, each AMPT
curve with finite thickness has been shifted a bit in time so that
it peaks at the same $t$ value as the corresponding dashed curve. 
Similar to Fig.~\ref{fig:emax}, one also sees in Fig.~\ref{fig:ampt0z}
that the increase of the maximum energy density$\emax$ with the
collision energy $\snn$ is much faster after one includes the finite time
duration of the initial energy production.

The analytical results of this study only address the energy density
at spacetime-rapidity $\eta_s=0$ in the center-of-mass frame of the
collision. Therefore the results for a realistic finite range of
spacetime-rapidity, e.g. $|\eta_s|<1/2$, would be somewhat different. 
Also, I have only addressed the energy density averaged over the full
transverse overlap area $\area$. Note that the transverse overlap area
at time before $\dt/2$ is smaller due to the partial
overlap of the two nuclei. To average over this partial overlap area, 
one may replace $\area$ in my solutions by $\area [1-(1-2t/\dt)^2]$
for $t \leq \dt/2$. This will enhance the energy density somewhat at
early times.
Also note that the finite duration of proper time in the initial
energy production has been considered in hydrodynamic models 
\cite{Okai:2017ofp,Shen:2017ruz}, where an energy source
term with a finite time duration can be introduced and my method 
can be applied to help describe the initial stage.

\section{Conclusions}
I have extended the Bjorken formula by including a 
time profile for the initial energy production due to the finite 
nuclear thickness. 
By considering a simple uniform as well as more
realistic time profiles, I have obtained analytical solutions 
of the formed energy density in the central spacetime-rapidity
region. 
These solutions approach the Bjorken formula at high collision
energies and/or at late times, but they are also valid at low energies
where the Bjorken formula breaks down. 
I then apply the solutions to central Au+Au collisions in the energy
range $\snn \in [4.84, 200]$ GeV. After taking into account the finite crossing
time, at lower energies where the crossing time is bigger, 
the maximum energy density achieved is much less sensitive
to the uncertainty of $\tauf$ and increases much faster with the
collision energy than the Bjorken formula.
At low energies, the energy density reaches a much lower
maximum value than the Bjorken energy density for the same formation
time $\tauf$, but the width of the time evolution of energy density is
much bigger. 
In addition, comparisons with the results from the string melting AMPT model
confirm the key features of the analytical solutions. 
The AMPT results also suggest that the effect of the
finite longitudinal width of the initial energy production on 
my analytical results is small. 
Therefore this extension provides a convenient tool to model the initial
energy production in relativistic heavy ion collisions, especially at 
low energies.

\section*{Acknowledgments}
I thank Miklos Gyulassy for careful reading of the
manuscript and helpful comments. This work is supported in part by
National Natural Science Foundation of China grant No. 11628508.

\end{document}